# Optimization identification of superdiffusion processes in biology: an algorithm for processing observational data and a self-similar solution of the kinetic equation

A.B. Kukushkin[1,2,3], A.A. Kulichenko[1], A.V. Sokolov[4]

[1]National Research Center "Kurchatov Institute", Moscow, 123182, Russian Federation
[2]National Research Nuclear University MEPhI, Moscow, 115409, Russian Federation
[3]Moscow Institute of Physics and Technology, Dolgoprudny, Moscow Region, 141700, Russian Federation
[4]Institute for Information Transmission Problems (Kharkevich Institute) of Russian Academy of Science, 127051 Moscow, Russian Federation

This work is an attempt to transfer to biology the methods developed in physics for formulating and solving the kinetic equations in which the kernel of the integral operator in spatial coordinates is slowly decreasing with increasing distance and belongs to the class of Lévy distributions. An algorithm is proposed for the reconstruction of the step-length probability density function (PDF) on a moderate number of trajectories of biological objects (migrants) and for the derivation of the Green's function of the corresponding integro-differential kinetic equation for the density of migrants in the entire space-time range, including the construction of an approximate self-similar solution. A wide class of time-dependent superdiffusion processes with a model power-law step-length PDF is considered, which corresponds to "Lévy walks with rests" for given values of the migrant's constant velocity and the average time $\tau$ of the migrant's stay between runs. The algorithm is tested within the framework of a synthetic diagnostics, consisting in the generation of artificial experimental data for trajectories of migrants and the subsequent reconstruction of the parameters of the step-length PDF and $\tau$. For different volumes of synthetic data, to obtain a general idea of the distributions under study (non-parametric case) and to evaluate the accuracy of recovering the parameters of the PDF (in the case of a parametric representation), the method of balanced identification is used. The approximate self-similar solution for the parameters of step-length PDF and $\tau$ is shown to provide reasonable accuracy of the space-time evolution of migrant's density.

## 1. Introduction

The study of superdiffusion processes (or, equivalently, nonlocal transport described by integral equations that are not reducible to differential equations or represented by equations in the so-called fractional derivatives) is expanding due to the breadth of the class of such phenomena in various fields of science (see, e.g., [1-5]). The slow decrease (with increasing distance) of the distribution function over the free path length of the carriers leads to the dominant contribution of long-free-path carriers. In the case of an instant response of the medium, such trajectories of carriers are called "Lévy flights" (the concept of "Levy flights" was introduced by B. Mandelbrot, see page IX in [1]), and in the case of finite velocity, such trajectories are called "Lévy walks" [6] (for more details see the review [2]). More precisely, taking into account the finite velocity of the carriers corresponds to a generalization of "Levy flights", which takes into account the effect of retardation in the Lévy flights and is called "Lévy walks with rests" (see Fig. 1 in [2]).

The phenomena of superdiffusion are actively studied in biology, see, e.g. review [7], popular review [8], monograph [9], section VI in [2], most recent example of Lévy flights [10]. In biology, in contrast to tasks in which the velocity of carriers (in biology, migrants) can be considered infinite, their finite velocity should always be taken into account. The finite velocity of the migrants makes the transfer dynamics substantially different from the case when the dominant characteristic time is only the "waiting time" (waiting time, for example, in the case of the transfer of resonant radiation in gases and plasma, is the radiative lifetime of an excited atom or ion).

The superdiffusion phenomena in biology, in contrast to most of those in physics, are characterized by the possibility of obtaining experimental data on individual trajectories of migrants, namely, animals or microorganisms in a wide range of sizes, including bacteria, people who agreed to participate in experiments on tracking their GPS movements on their phones). This allows one to restore the probability density function (PDF) of migrant's free path (step-length PDF) and the distribution function of the migrant's waiting time (i.e. the time of stay (rest) between runs) using a limited set of trajectories of the studied objects. The knowledge of these distribution functions is widely used to estimate the average distance by which a migrant moves for a given time or, equivalently, the average time for which a migrant moves by a given distance (the introduction of such concepts assumes the uniformity and stationarity of the environment in which the migrant moves, otherwise the indicated dependences will also be functions of the starting moments of time and the initial coordinates). The data processing and analysis of the step-length PDF for belonging to the class of long-tailed distributions (such as the Lévy distribution) is rich in examples, while the answer to the question whether animals obey Lévy statistics remains controversial (see, for example, the beginning of Section VII in [2]).

However, we could not find examples of the developed formalism of kinetic equations for the Green's function, which describes, in its very meaning, the space-time dependence of the density of migrants who begin to move at a given moment from a given point in space. Moreover, there is no practice of obtaining an explicit dependence of the Green's function on space coordinates and time. Therefore, it makes sense to try to transfer to biology the methods developed in physics for formulating and solving kinetic equations in which the kernels of the integral operators in spatial coordinates are slowly decreasing at large distances and belong to the class of Lévy distributions. Such an attempt is appropriate to be carried out in the framework of synthetic diagnostics, consisting in the generation of artificial experimental data for migrant trajectories and the subsequent reconstruction of the parameters of the step-length PDF and the average waiting time $\tau$ of the migrants between runs.

Recall that in the case of normal (or ordinary) diffusion, defined as a Brownian motion described by a Fokker-Planck differential equation, the Green's function is a Gaussian whose argument defines the dynamics of the propagation front, $r_{fr}(t) \sim (Dt)^\beta$, where $\beta = ½$, D is the diffusion coefficient. This law is violated in a wide class of phenomena, where the step length PDF decreases slowly with increasing distance, according to the power law. This leads to the divergence of the diffusion coefficient formally determined from the dispersion of the step-length PDF and to the exponent $\beta > ½$ in the front propagation law, $r_{fr}(t) \sim (t)^\beta$. It is this case which is called superdiffusion transfer.

In a number of physical problems, the phenomenon of superdiffusion is usually called nonlocal transport, which is described by an integral, in spatial variables, equation that is not reducible to a differential equation: for example, the Biberman-Holstein equation [11, 12], which describes the spatiotemporal evolution of the density of excited atoms or ions, caused by radiation transfer in spectral lines in plasma and gases [13, 14]; excitation transfer by nonequilibrium resonant phonons in condensed matter [15]; (non-stationary) heat transfer by longitudinal (electron Bernstein) plasma waves [16]; photoinduced transport of minor carriers in semiconductors (namely, hole transport in n-type semiconductors associated with photon absorption) [17]. By the example of the Biberman-Holstein model, in which resonant scattering of photons by an atom or ion occurs with a complete frequency redistribution in the act of absorption and reemission of a photon, the main property of superdiffusion transport is well traced: rare long-distance flights of photons ("jumps") that correspond to emission and absorption in the "wings" of the spectral line prevail over the contribution of frequent close displacements that cause the diffusion (Brownian) motion of medium excitation and correspond to emission and absorption in center of the spectral line. It was shown in [18] that distant jumps caused by slowly falling (for example, power-law) wings of the integral operator (i.e., step-length PDF) in the transport equation are Lévy flights. The dominant contribution of photons with a long free path to radiative transfer in spectral lines was already shown in [19, 20].

For time-dependent superdiffusion transport by Lévy flights, it was recently shown that for a wide class of transport phenomena on a uniform background, an approximate self-similar solution for the Green's function can be obtained [21-26]. These solutions were constructed using scaling laws for the perturbation propagation front (i.e., the time dependence of the average distance of the superdiffusion carrier from the point instant source) and asymptotic expressions for the Green's function far behind and far ahead of the front. The validity of the proposed self-similar solutions was shown by comparing them with exact numerical solutions: in the one-dimensional case, the solutions of the transport equations for a simple slowly decaying step-length PDF with various power exponents; in the three-dimensional case, the solutions of the Biberman-Holstein resonance radiative transfer equation for various spectral line shapes (Doppler, Lorentz, Voigt, and Holtsmark).

The method [21] of approximate self-similar solutions for the Green's function of time-dependent superdiffusion (nonlocal) transport equations was generalized to the case of finite fixed velocity of carriers in [27] for the problems of one-dimensional perturbation transfer in a homogeneous medium for a simple step-length PDF with a power-law decay at large distances. The general [28] and approximate self-similar solutions [27, 29, 30] were found for an arbitrary superdiffusion step-length PDF and, using the numerical calculation of the general solution, the accuracy of the self-similar solution was verified for a specific power law of the step-length PDF, which, for example, corresponds to the Lorentz shape of the wings of the atomic spectral line shape for photon emission.

In this paper, we consider a wide class of non-stationary superdiffusion processes with a model power-law step-length PDF corresponding to "Levy walks with rests" for given values of the constant velocity of migrants and the average waiting time $\tau$ between migrant's runs. In Section 2, a two-dimensional, in spatial coordinate, integro-differential kinetic equation for the density of migrants is proposed. Next, an algorithm is proposed for reconstructing the basic parameters of the kinetic equation, including the probability density function of migrant's free path (i.e. the step-length PDF), using a moderate number of migrant trajectories (Section 3). Next, the algorithm is tested within the framework of synthetic diagnostics, consisting in the generation of artificial experimental data for migrant trajectories on a homogeneous plane and the subsequent reconstruction of the parameters of the step-length PDF and $\tau$ (Section 4). In Section 5, we construct an approximate self-similar solution for the Green's function of the indicated kinetic equation. The conclusions are made in Section 6.

**2. Kinetic equation for the density of migrants and their trajectories**

Consider the process of superdiffusion transfer, which is the migration under the following conditions:
- the migrant's trajectory consists of runs and stops,
- migration characteristics do not depend on time and coordinates on the plane (i.e. external conditions for migrants are homogeneous and stationary),
- the module of the velocity of motion, c, between stopping points is constant in time,
- the distribution in the direction of the velocity vector for motion after stopping is isotropic in the plane,
- the probability density function (step-length PDF) for a free path over a distance ρ has the form

$$W_{step}(\rho) = \frac{\gamma \kappa_0}{(1 + \kappa_0 \rho)^{\gamma+1}}, \quad 0 < \gamma < 2,$$

$$\int_0^\infty W_{step}(\rho')d\rho' = 1$$

(1)

where the constant $1/k_0$ is the characteristic length of free path. For two-dimensional motion, the step-length PDF (1) corresponds to the following PDF for stopping at a point $r$ after starting at a point $r_1$ (or vice versa):

$$W(\rho) = \frac{W_{step}(\rho)}{2\pi\rho}$$

$$\rho = |r - r_1|, \, 0 < \gamma < 2,$$

$$\int W(|r - r_1|)d^2 r = 1$$

(2)

(recall that for the PDF (2) the mean square displacement is infinite),
- the probability distribution over waiting time (i.e. the waiting time PDF) has the form

$$U(t) = \frac{1}{\tau}\exp\left(-\frac{t}{\tau}\right), \quad \int_0^\infty U(t)dt = 1,$$

(3)

where τ is the average lifetime at the stopping point (or, equivalently, the average waiting time).

Then the equation for the Green's function of the density of standing migrants, $f(r,t)$, at the point $r = \{x, y\}$ at time t has the form:

$$\frac{\partial f(r,t)}{\partial t} = -\left(\frac{1}{\tau} + \sigma_{sink}\right)f(r,t) + \frac{1}{\tau}\int d^2 r_1 W(|r - r_1|)\theta\left(t - \frac{|r - r_1|}{c}\right)f\left(r_1, t - \frac{|r - r_1|}{c}\right)$$
$$+ \delta(r)\delta(t)$$

(4)

where θ is the Heaviside unit function, $\sigma_{sink}$ is the average inverse time of the disappearance (sink) of migrants. This equation is a two-dimensional analog of equation (1) in [28], its derivation for a similar the process — the transfer of resonant radiation in gases and plasma with the finite photon velocity taken into account — can be found in Section 2 in [28], and the derivation without accounting for retardation is given in the Appendix in [21]). The initial condition in (4) corresponds to a single standing migrant at the origin.

The general solution of equation (4) obtained in the way similar to [28] has the following form [30]:

$$f(r,t,R_c) = \frac{1}{(2\pi)^2 i} \int_0^{+\infty} dp\, p\, J_0(p\kappa_0 r) \int_{+0-i\infty}^{+0+i\infty} \frac{e^{st/\tau}\, ds}{s + 1 + \sigma_{sink}\tau - \gamma \int_0^{+\infty} \frac{du\, e^{-su/R_c}}{(1+u)^{\gamma+1}} J_0(pu)} \quad (5)$$

where $J_0$ is the Bessel function, and the parameter

$$R_c = c\tau\kappa_0 \quad (6)$$

is a ratio of the average waiting time to the average flight time (average time at motion) of a migrant.

The results of numerical calculations of the general solution (5) are given in Section 5.

Equation (4) corresponds to the following type of migrant's trajectories. At a given point (we take it as the origin), a migrant stands time t, which is a random variable determined by the waiting time PDF $U(t)$ (3). At a specified time moment, the migrant begins to move with equal probability in any direction (i.e., uniformly in angle in the plane of motion) along a straight path with a constant velocity c. The migrant goes along this direct path, the length of which is a random variable described by the step-length PDF $W_{step}(\rho)$ (1). At the stopping point, the migrant stands the time determined by distribution (3) and again moves equally likely in direction, going a distance determined by distribution (1). The process is repeated over and over again.

Figure 1 shows two examples of the trajectory of the indicated type. One can see the characteristic feature of trajectories with long steps (Levy flights), which connect the clusters of (mostly multiple) short steps which form a walk of the Brownian type. Artificial (synthetic) experimental data for $\gamma = 1.5$, $\kappa_0 = 1$, $\tau = 1$, c =10, containing the migrant trajectories on the (uniform) plane for 100 migrants for a period of dimensionless time t = 1000, were prepared using the pseudo-random number generator. The corresponding table of state changes (from rest to motion and back to the rest) containing more than 150 thousand rows can be found at
https://drive.google.com/file/d/1BuQQgtCBdC5VbVrLMvgb7U-VHIGt7Dy4/view?usp=sharing

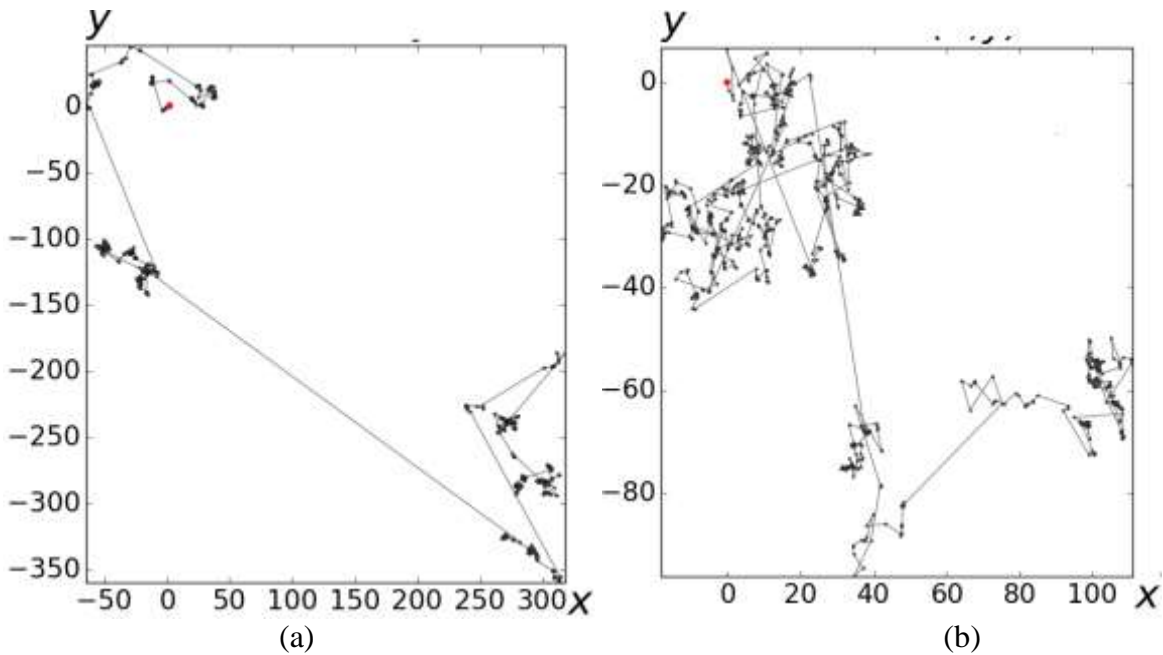

(a)         (b)

Fig. 1. Typical trajectory of a migrant who stood at the origin at zero time (red dot), for step-length PDF (1) with γ = 1.5, for a period of dimensionless time 1000 (i.e. divided by τ), depending on dimensionless coordinates (i.e. multiplied by $k_0$). The points of stops (breakpoints) are indicated by dots.

An example of the space-time evolution of the density of migrants obtained with the Monte-Carlo modeling is given in Figure 2. A comparison of the results of such a modeling with the exact and approximate self-similar solutions of the kinetic equation is presented in Sec. 5.

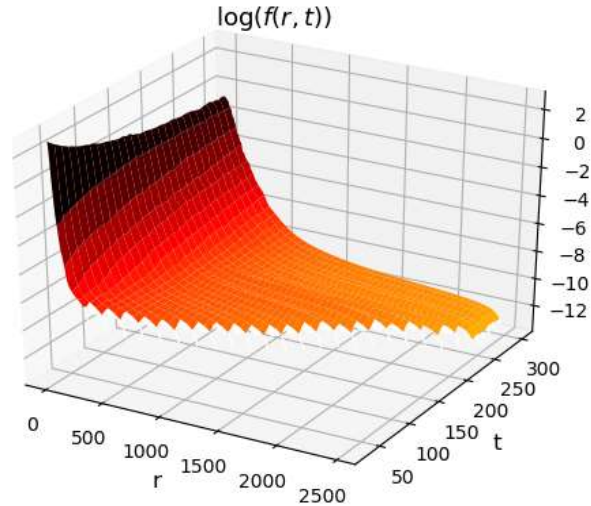

Figure 2. Space-time evolution of the density of standing migrants obtained with the Monte-Carlo modeling for $10^4$ standing migrants at r=0, t=0 for $\gamma = 1.5$, $\kappa_0 = 1$, $\tau = 1, c = 10$.

## 3. An algorithm for reconstructing the basic parameters of migrant's motion from a moderate number of trajectories

Here we present an algorithm for reconstructing the basic parameters of migrant's motion, which are needed for formulating the kinetic equation, namely the parameters of the step-length PDF (1) and the waiting time PDF (3). The algorithm consists in the identification, using the balanced identification method [30, 31], of the sought-for distributions, using a limited set of migrant trajectories. Here we use the generated array of migrant trajectories described in previous section.

In the general case, all parameters (constants) that determine the trajectories of migrants (and determine the solution of the transfer equation (4)) may be the sought-for parameters, namely:
- characteristic length of free path, namely the constant $1/k_0$ (it is the constant that the spatial coordinates are expressed in units of it in the dimensionless description),
- a parameter that describes the dispersion of the step-length PDF, namely, the exponent γ in (1),
- average lifetime of a migrant at the stopping point (or capture), i.e. the average waiting time, τ,
- the delay parameter, $R_c = c\tau\kappa_0$, which is the ratio of the average waiting time to the average flight time.

Below, in accordance with (4), we restrict ourselves to the task of restoring the first three parameters without introducing stochasticity in the value of the migrant's velocity. Also we consider only the case of $\sigma_{sink} = 0$ (no sink of migrants).

In the general case (without being tied to a specific parameterization), the algorithm for identifying the PDF consists in a regularized fitting of the corresponding distribution function (namely, the integral of the PDF over space coordinate). The task is to choose the optimal compromise between the proximity of distribution function to the original data and the simplicity of the PDF, expressed in terms of function's "curvature". Let us consider the statement of the problem and its solution for $W_{step}(\rho)$ (up to the notation, a similar formulation is possible for the function $U(t)$).

Denote the source data obtained from the trajectory array as a dataset:

$$\Omega: \quad \{F_k, \rho_k\}, k \in K, K = 1..k_{max}. \tag{7}$$

These data are shown in Figure 2 (top figure, $F_k$ data points).
The task is to find the functions $F$ and $W_{step}$, related by the equation

$$F(\rho) = \int_0^\rho W_{step}(\rho')d\rho', \tag{8}$$

using the dataset $\Omega$. To formulate and regularize the problem, we use the balanced identification method [31, 32], which is based on minimizing a functional that depends on the regularization coefficient $\alpha$. So, the solutions of the identification problem are the functions $F$ and $W_{step}$, related by equation (8) and providing a minimum of the functional:

$$Q(F, W_{step}, K, \alpha) = \frac{1}{|K|}\sum_{k \in K}(F_k - F(\rho_k))^2 + \alpha \int_0^{\rho_{max}} \left(\frac{d^2 W_{step}}{d\rho^2}\right)^2 d\rho' \to \min_{F, W_{step}}, \tag{9}$$

where the first term is a measure of the model's compliance with the measurements (standard deviation from data), the second is the regularizing additive, which is a measure of model complexity, defined as the curvature of the function $W_{step}$ (the square of the second derivative).

Here, $\alpha$ is a regularization parameter. Different values of it determine models with a different ratio of deviation of the function $F$ from the data (7) and the complexity of the function $W_{step}$. Figure 3 shows three examples of numerical solutions to problem (7) - (9) for various values of $\alpha$. If $\alpha$ is too large, the second term in (9) suppresses the first (Fig. 3A). The resulting model is too simple: the normalized (i.e. divided by the standard deviation of the dataset) root mean-squared error (rmse), and the normalized cross-validation error ($\sigma_{cv}$) are too large. The other extreme, too small $\alpha$, leads to the dominance of the first term (Fig. 3C) and, as a result, although the standard deviation is small, the step-length PDF is too chaotic, as evidenced by the value of the cross validation error. Figure 3B shows an optimally balanced solution that implements the optimal trade-off between the proximity of the model to the data and the simplicity of the model. The optimal value of $\alpha$ is selected by minimizing the mean square error of the cross-validation $\sigma_{cv}$ [31, 32].

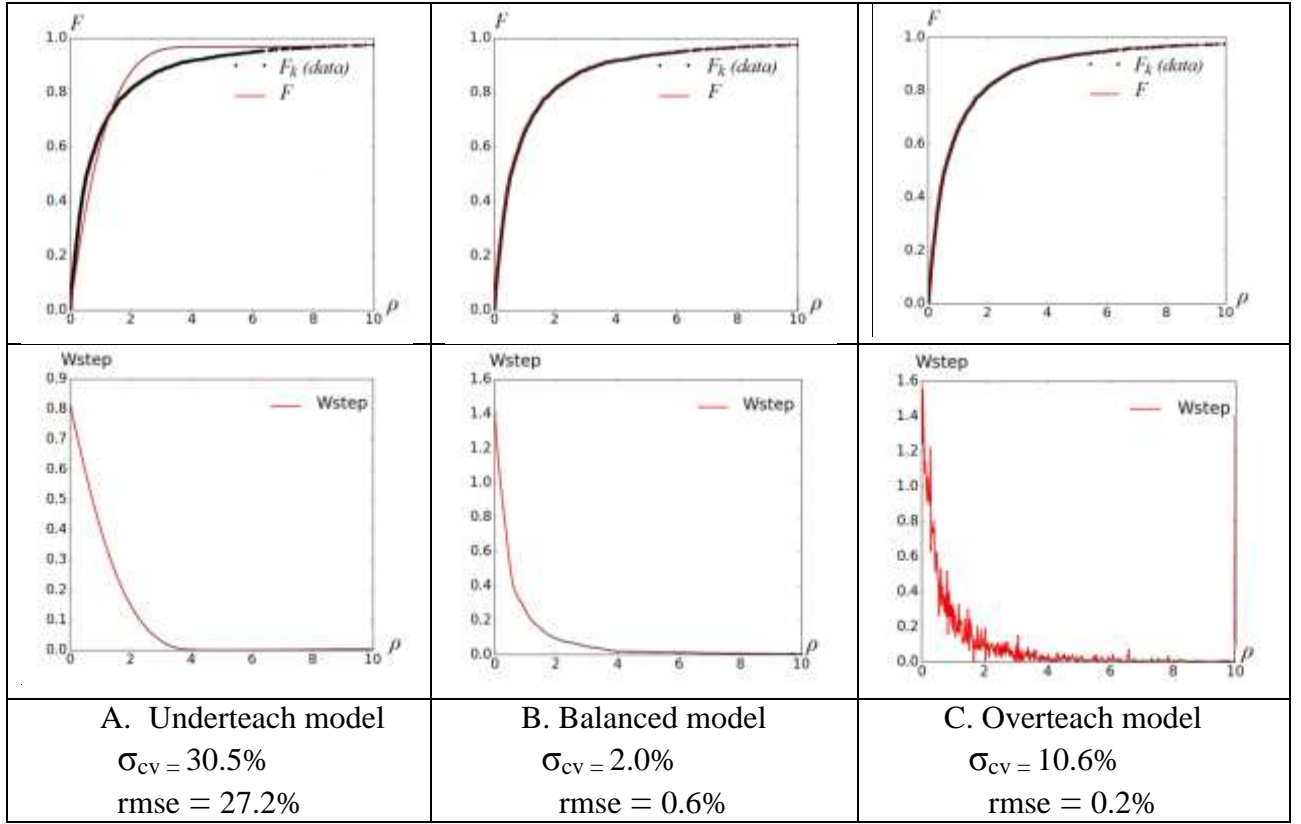

Fig. 3. Various versions of data approximation: the lower graphs are the step-length PDF in Eq. (1), $W_{step}$, the upper graphs are the corresponding distribution functions F in Eq. (8).

Note that for cross-validation we used the subdivision of the dataset (7) into seven approximately equal parts in sequence. In addition, the fields of about 5% of the sample size were discarded from the training sample (on each side). The use of such a complex procedure is explained by the connectedness (not randomness) of the neighboring values of the set of experimental data (in our case, synthetic data for trajectories).

## 4. Testing the algorithm in the framework of synthetic diagnostics

We will test the algorithm for reconstructing the basic parameters of migrant's motion from a moderate number of trajectories, described in previous section, in the framework of synthetic diagnostics, which consists in restoring the distribution functions (both in parametric representation and in general form) based on artificial (synthetic) experimental data for migrant trajectories on a homogeneous plane, which are described in Section 2. Testing was conducted for data samples of various sizes (Table 1).

The balanced identification method, in addition to identifying the functions, allows to get a fairly reliable estimate of the data modeling error, namely, the mean square error of cross-validation ($\sigma_{cv}$). Such errors for the distribution functions of free path ($F_{Wstep}$ from Eq. (8)) and waiting time ($F_U$, similar distribution function for the PDF (3)) are shown in Table 1. An analysis of the results shows the efficiency of the method used to obtain nonparametric distribution functions. However, it is not possible to obtain similar estimates for the PDFs ($W_{step}$ and U). Therefore, the question of how the restoration error of these PDFs depends on the data size is investigated experimentally. To do this, we calculate the PDFs using the samples of different size. The calculation results show that the sample's size of 3500 tests (i.e. measurements of step length or of waiting time) seems to be sufficient and, with a further increase in the amount of data, the functions change insignificantly. Figure 4 shows the PDFs obtained for various amounts of the initial data (700, 3500, and 7000 tests).

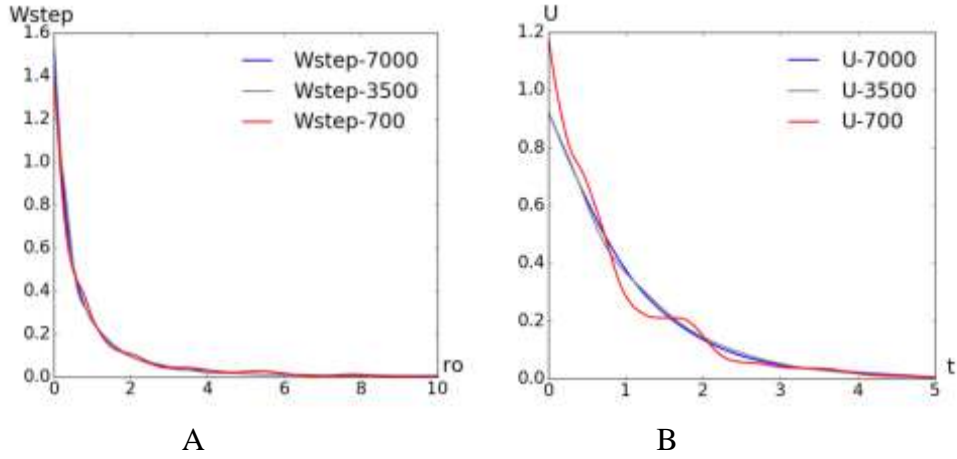

A                          B

Fig. 4. Probability density functions restored from synthetic experimental data of Sec. 2 in the case of non-parametric representation of the sought-for functions: step-length PDF (A) and waiting time PDF (B). The number of measurements of the step length and waiting time in the input data are shown in the legend.

The parametric representation of the distributions significantly reduces the freedom of choice. Therefore, to identify the parameters, the usual minimization of the mean square error (rmse) is sufficient. However, we again use the balanced identification procedure to obtain estimates of the error $\sigma_{cv}$.

In particular, to evaluate the parameters of $W_{step}$ (1), it is necessary to substitute into the identification criterion (9) the PDF (1) and the corresponding parametric distribution function

$$F_{Wstep}(\rho) = 1 - \frac{1}{(1+\kappa_0\rho)^\gamma}, \tag{10}$$

which can be easily obtained by integration (8). Identification criterion for $U(t)$ (3) is formulated in a similar way.

The results of identification are shown in Table 1.

Tab. 1. The restored parameters of distributions and estimation of the modeling error $\sigma_{cv}$ for various data set sizes, for $\kappa_0 = 1$, $\gamma=1.5$, $\tau=1$.

| Size of data sample (number of tests) | Time | Non-parametric | | Parametric $W_{step}$ | | | Parametric $U$ | |
|---|---|---|---|---|---|---|---|---|
| | | $\sigma_{cv}$ ($F_{Wstep}$) | $\sigma_{cv}$ ($F_U$) | $\gamma$ | $\kappa_0$ | $\sigma_{cv}$ | $\tau$ | $\sigma_{cv}$ |
| 700 | 16 | 3.41 | 1.51 | 1.359 | 1.024 | 2.26 | 1.107 | 3.91 |
| 3500 | 39 | 1.97 | 1.39 | 1.499 | 1.044 | 2.24 | 0.968 | 1.45 |
| 7000 | 86 | 0.38 | 1.07 | 1.453 | 1.079 | 1.38 | 0.980 | 2.10 |
| 35000 | 509 | 0.17 | 0.87 | 1.454 | 1.047 | 0.37 | 0.997 | 0.87 |
| 70000 | 946 | 0.11 | 0.30 | 1.503 | 0.999 | 0.48 | 0.996 | 0.60 |

Table 1 shows that for the sample's size of 3500 tests or more, the accuracy of the reconstruction of the parameters is satisfactory. Note that the size of 70,000 measurements approximately corresponds to the volume contained in the synthetic data for trajectories of 100 migrants during dimensionless time t = 946 for $\tau =1$ (see Section 2).

## 5. Exact and approximate self-similar solutions of the kinetic equation for the Green's function

In this Section, we present numerical solutions of eq. (4) for the case $\gamma = 1.5$, construct an approximate automodel solution of eq. (4), and analyze the accuracy of self-similarity (automodelity).

The method of constructing approximate self-similar solutions [21-26] was generalized in [27], taking into account the finite velocity of the carriers. According to the specified method, an approximate self-similar solution has the form (in (11)-(15) in dimensionless variables of time, in units of $\tau$, and space, in units of $1/k_0$):

$$f_{auto}(r,t,R_c,\gamma) \equiv F_{auto}(r,t,R_c,\gamma,g(r,t,R_c,\gamma)) = \left(t - \frac{r}{R_c}g(s)\right)W(rg(s))\theta\left(t - \frac{r}{R_c}g(s)\right)$$
(11)

$$s \equiv \rho_{fr}(t,R_c)/r \tag{12}$$

where

$$g(s) = \begin{cases} 1, & s = s_{min} = \rho_{fr}(t,R_c)/(R_c t), \\ s, & s \gg s_{min} \end{cases}$$

and the propagation front of perturbations $\rho_{fr}(t,R_c)$ (i.e., the time dependence of the average distance of the migrant from the source) is defined as (8) in [21],

$$\left(t - \frac{\rho_{fr}}{R_c}\right)W(\rho_{fr})\theta\left(t - \frac{\rho_{fr}}{R_c}\right) = f(0,t,R_c)$$
(13)

To estimate the accuracy of the proposed approximate solution (11), we introduce the function $Q(r,t,R_c)$ (cf. (9) in [21])

$$\left(t - \frac{r}{R_c}Q(r,t,R_c)\right)W(rQ(r,t,R_c))\theta\left(t - \frac{r}{R_c}Q(r,t,R_c)\right) = f(r,t,R_c)$$
(14)

To verify the accuracy of the indicated automodelity, it should be shown that the function $Q_1$

$$Q_1(s,t,R_c) = Q(\rho_{fr}(t,R_c)/s,t,R_c) \approx g(s,R_c)$$
(15)

has a weak dependence on time $t$. The dependence $Q_1(s,t,R_c)$ on the variable $s$ at various moments of time at $\gamma = 1.5$ for $R_c = 1$ and $R_c = 10$ was shown in Fig. 4(c) in [30] for the range of time values is in the interval $t = 30 - 3000$.

The accuracy of (14) was verified by calculating the maximum relative deviation of $Q_1(s,t,R_c)$ from $Q_1(s,t = 3000,R_c)$ for different time moments at $\gamma = 1.5$ for $R_c = 1$ and $R_c = 10$. The comparative value of the function was chosen at the greatest moment of time, since with increasing $t$ the automodelity function is less dependent on time. The main difference between the dependences $Q_1(s,t,R_c)$ and $Q_1(s,t = 3000,R_c)$ is observed near $s = s_{min}$, while the maximum relative deviation is about 30% for $R_c = 1$ and 70% for $R_c = 10$.

Further, using the available self-similarity functions $Q_1(s,t,R_c)$, one can construct approximate self-similar solution (11) and compare it with general solution (5). For comparison, we may use

$Q_1(s, t = 100, R_c)$ or $Q_1(s, t = 3000, R_c)$ as the main self-similarity function (note that in the second case we have the weakest time dependence).

For testing the accuracy of self-similar solutions (11), including the effect of the deviation of the parameters of the trajectories, $\tau, \kappa_0, \gamma$, reconstructed from certain volume of available observational data, from their true values (see sec. 4), we introduce the following functions for analyzing the case of true values $c = 10, \tau = 1, \kappa_0 = 1, \gamma = 1.5$:

$$D(\frac{r}{ct}, t, \mu, \gamma) = \frac{F_{auto}(r,t,R_c,\gamma,Q_1(s,t_{base},R_c))}{f_{exact}(r,t,\ c=10,\tau=1,\kappa_0=1,\gamma=1.5)} \qquad \mu = \{c, \tau, \kappa_0\}, \tag{16}$$

$$\Delta(\frac{r}{ct}, t, \mu, \gamma) = \frac{|F_{auto}(r,t,R_c,\gamma,Q_1(s,t_{base},R_c)) - f_{exact}(r,t,c=10,\tau=1,\kappa_0=1,\gamma=1.5)|}{f_{exact}(r=0,t,\ c=10,\tau=1,\kappa_0=1,\gamma=1.5)} \tag{17}$$

where $t_{base}$ corresponds to the choice of the self-similarity function (15) at a certain time moment to describe the solution for arbitrary time, $R_c$ is given by Eq. (6).

Figures 5 and 6 show, respectively, the functions D (16) and $\Delta$ (17) for $\gamma = 1.5$, $R_c = 10$ for five various time instants.

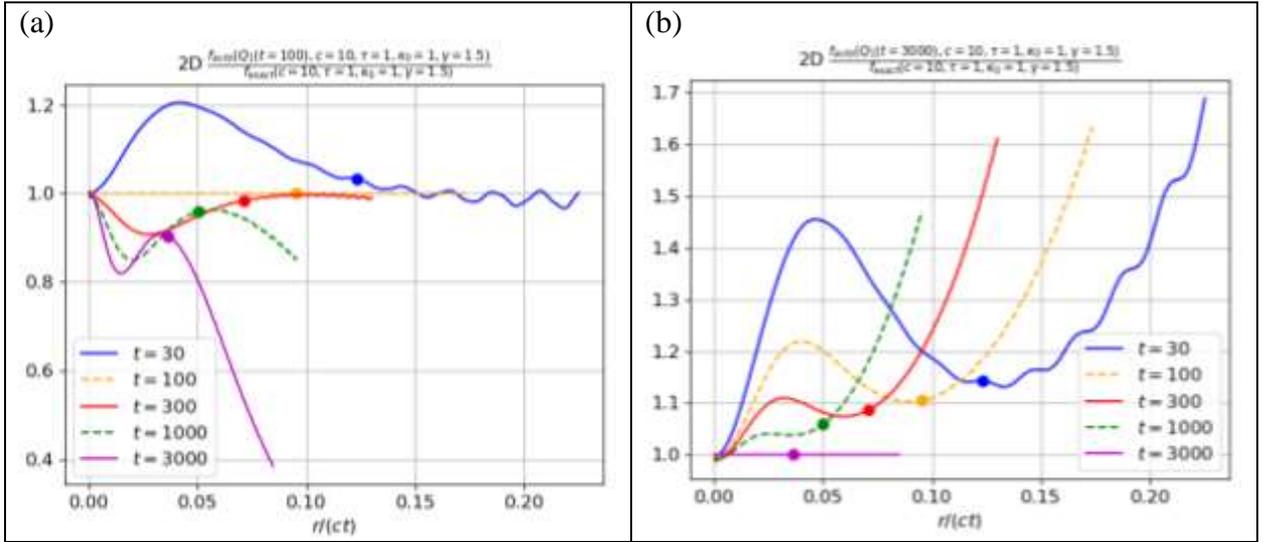

Figure 5. The ratio (16) for $\mu = \{\ c = 10, \tau = 1, \kappa_0 = 1\}, \gamma = 1.5$, and (a) $t_{base} = 100$, (b) $t_{base} = 3000$. Radial coordinate for each curve is limited to the region where the exact solution is not less than $10^{-3}$ of the value of this solution at $r = 0$. Bold dots indicate points where $f_{exact}$ is 1% of $f_{exact}(r = 0)$.

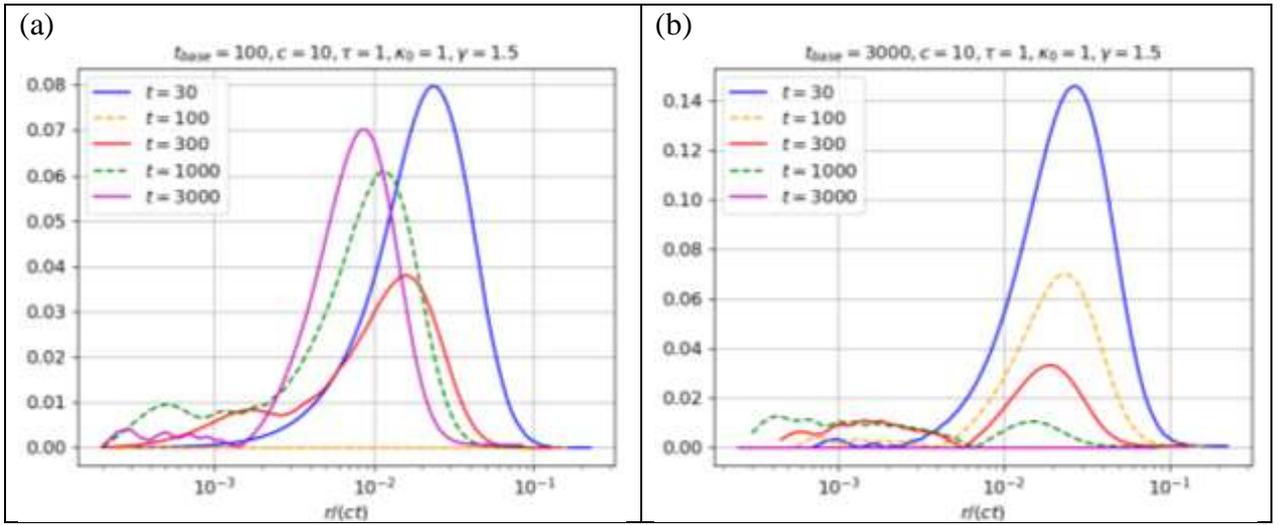

Figure 6.The ratio (17) for $\mu = \{c = 10, \tau = 1, \kappa_0 = 1\}, \gamma = 1.5$, and (a) $t_{base} = 100$, (b) $t_{base} = 3000$. Radial coordinate for each curve is limited to the region where the exact solution is not less than $10^{-3}$ of the value of this solution at $r = 0$.

Figures 5 and 6 show a good agreement between the approximate automodel solution (11) and the general solution (5) calculated numerically.

## 6. Test of approximate solution for recovered parameters of migration

The final step of the present work is the test of the ability of the approximate self-similar solution, obtained for the parameters of step-length PDF and τ reconstructed from a moderate number of migrant's trajectories, to provide reasonable accuracy of the space-time evolution of migrant's density. To try the capabilities of the self-similar solution we analyze the case of the parameters of step-length PDF and τ reconstructed from the minimal volume of data, namely those shown in the first and second lines in Table 1.

The self-similarity function for the above mentioned case is shown in Figure 6.

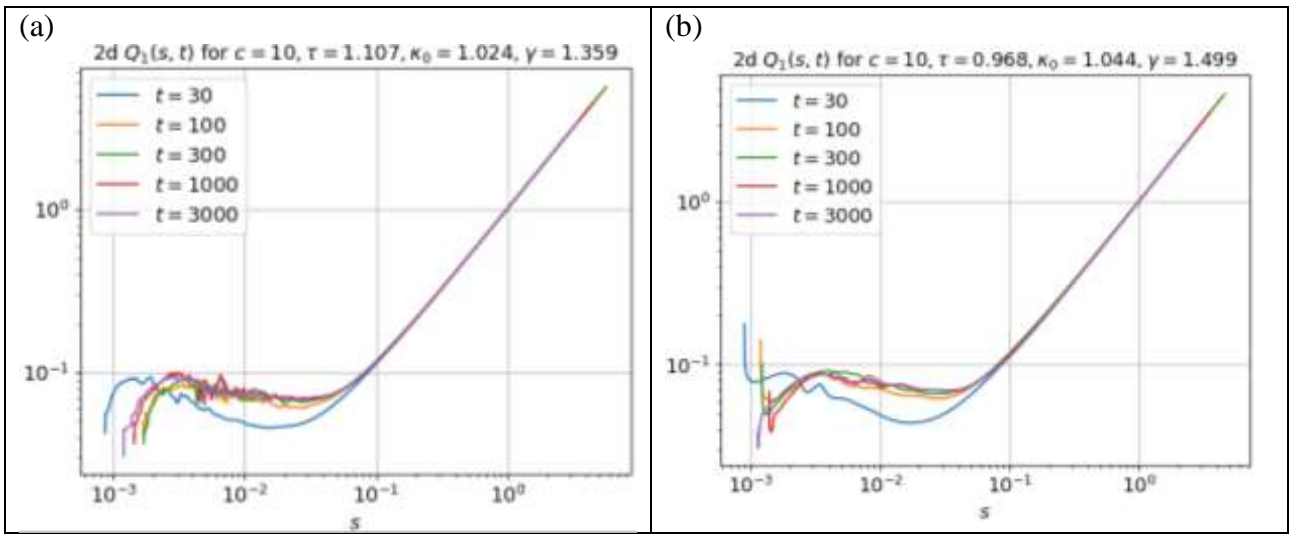

Figure 6. Self-similarity functions recovered from comparing the self-similar solution with exact solution for various times and c = 10, (a) $\tau = 1.107$, $\kappa_0 = 1.024$, $\gamma = 1.359$ (cf. first line in Table 1) and (b) $\tau = 0.968$, $\kappa_0 = 1.044$, $\gamma = 1.499$ (cf. second line in Table 1).

It is seen that starting from dimensionless time $t = 100$ the degree of self-similarity is high enough. Taking the minimal time (and consequently, spending less time in generating the exact solution) we arrive at the following results for the approximate self-similar solution in the entire interval $t = 30 - 3000$, shown in Figures 7 and 8.

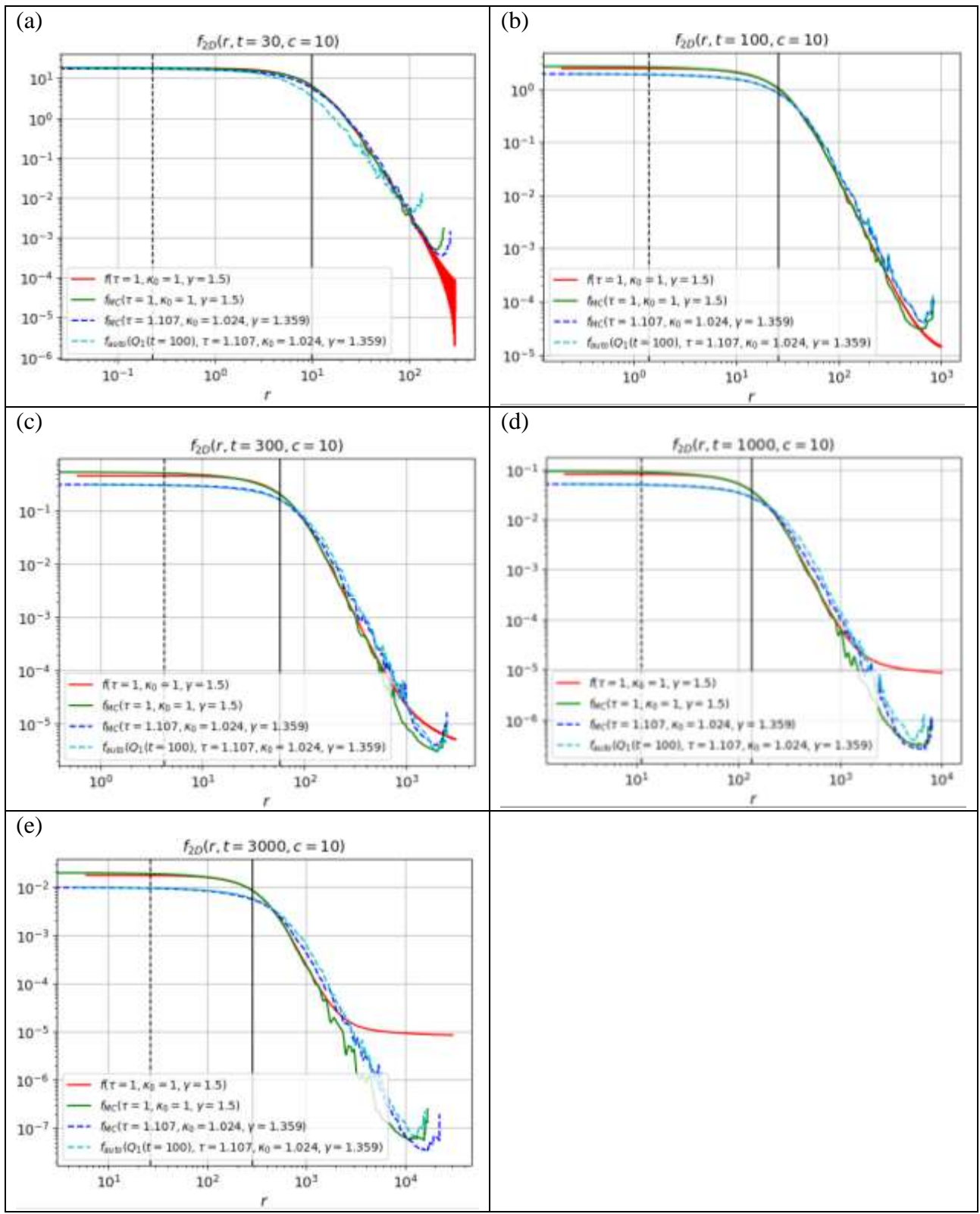

Figure 7. Comparison of various solutions to the problem at different points in time. Here, the red curves correspond to numerical calculation of general solution (5) for the true values of parameters: $c = 10$, $\tau = 1$, $\kappa_0 = 1$, $\gamma = 1.5$. The green and blue dashed curves correspond to the Monte Carlo calculation for, respectively, the true parameters

and reconstructed ones, $c = 10$, $\tau = 1.107$, $\kappa_0 = 1.024$, $\gamma = 1.359$, taken from Table 1, first line. The blue dashed curve shows an approximate self-similar solution obtained with the help of the self-similarity function at time $t_{base} = 100$ for reconstructed parameters. The vertical solid and dashed lines show the position of the propagation front (13) for, respectively, true and reconstructed parameters.

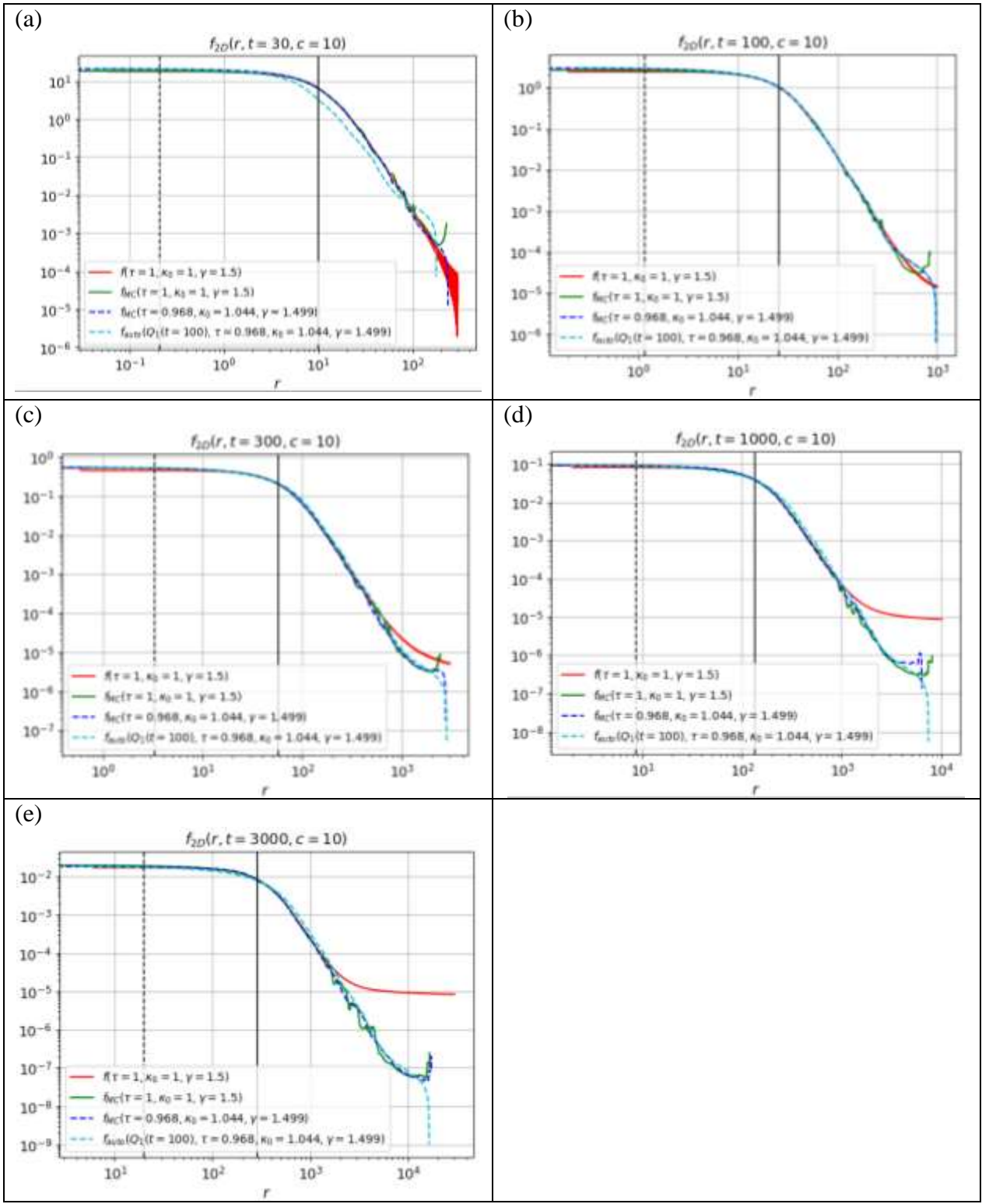

Figure 8. The same as in Fig. 7 but for reconstructed parameters, $\tau = 0.968$, $\kappa_0 = 1.044$, $\gamma = 1.499$, taken from the second line in Table 1.

Taking the self-similarity function recovered from exact solution at time $t_{base} = 3000$ for reconstructed parameters does not change the results of comparison in the form of Figures 7 and 8. This is because the self-similar solutions are close enough as shown below in Figure 9.

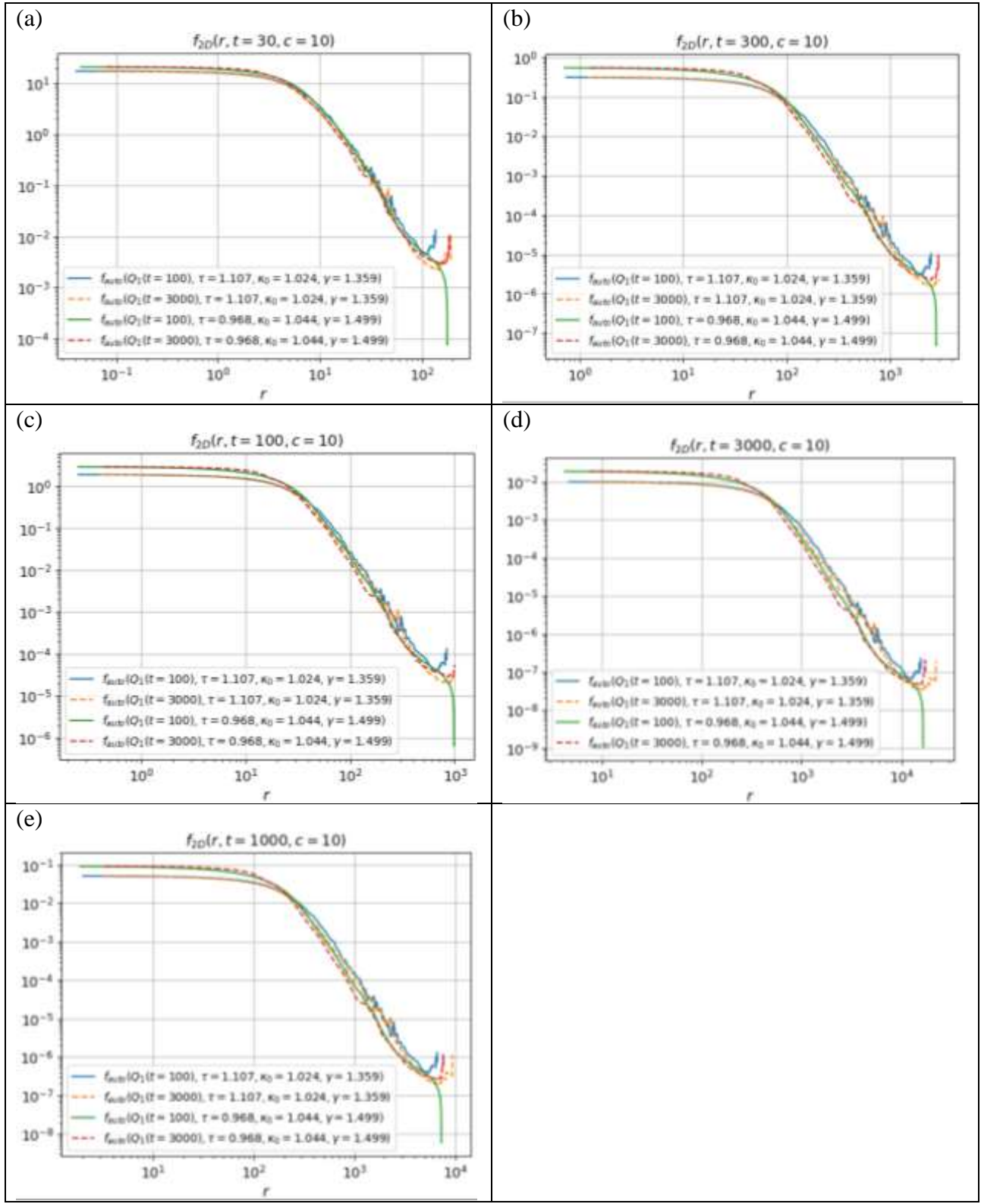

Figure 9. Comparison of self-similar solutions obtained with the help of self-similarity function recovered from exact solution for reconstructed parameters (Table 1, first line) at time $t_{base} = 100$ (blue) and $t_{base} = 3000$ (orange dashed) and for reconstructed parameters (Table 1, second line) at time $t_{base} = 100$ (green) and $t_{base} = 3000$ (red dashed).

The accuracy of obtained approximate self-similar solutions for two different cases of reconstructed parameters of migrant's trajectories, namely, first and second line in Table 1, is analyzed in Figures 10-13.

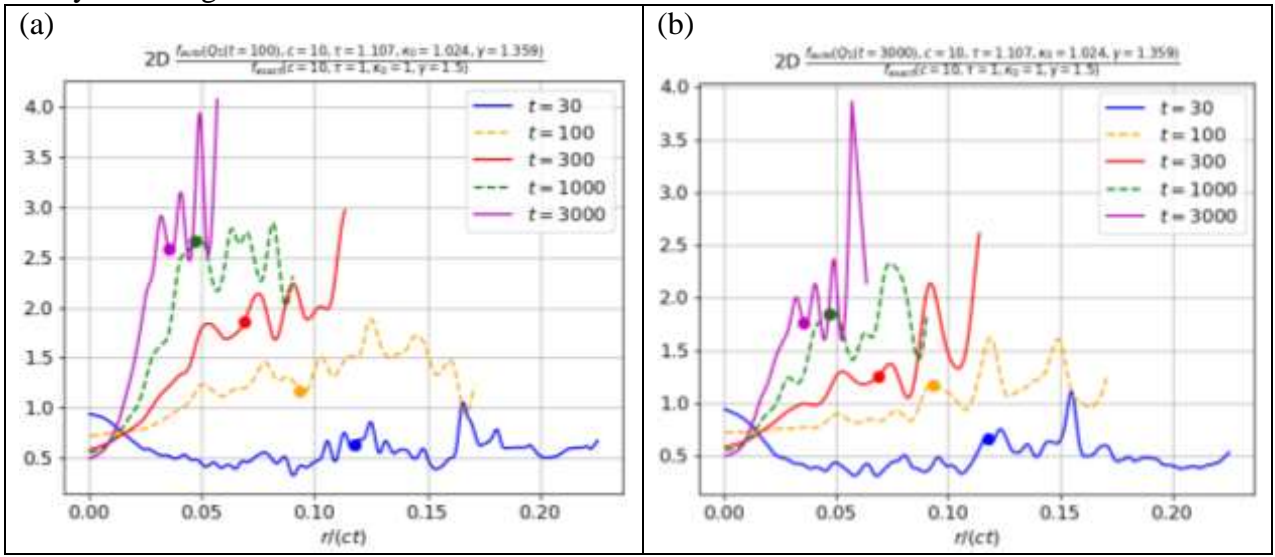

Figure 10. The ratio (16) of the self-similar solution to exact solution, when trajectory parameters in the self-similar solution are taken as reconstructed (Table 1, first line) and (a) $t_{base} = 100$, (b) $t_{base} = 3000$. Here, the radial coordinate for each curve is limited to the region where the exact solution is not less than $10^{-3}$ of the value of this solution at $r = 0$. Bold dots indicate point where $f_{exact}$ is 1% of $f_{exact}(r = 0)$.

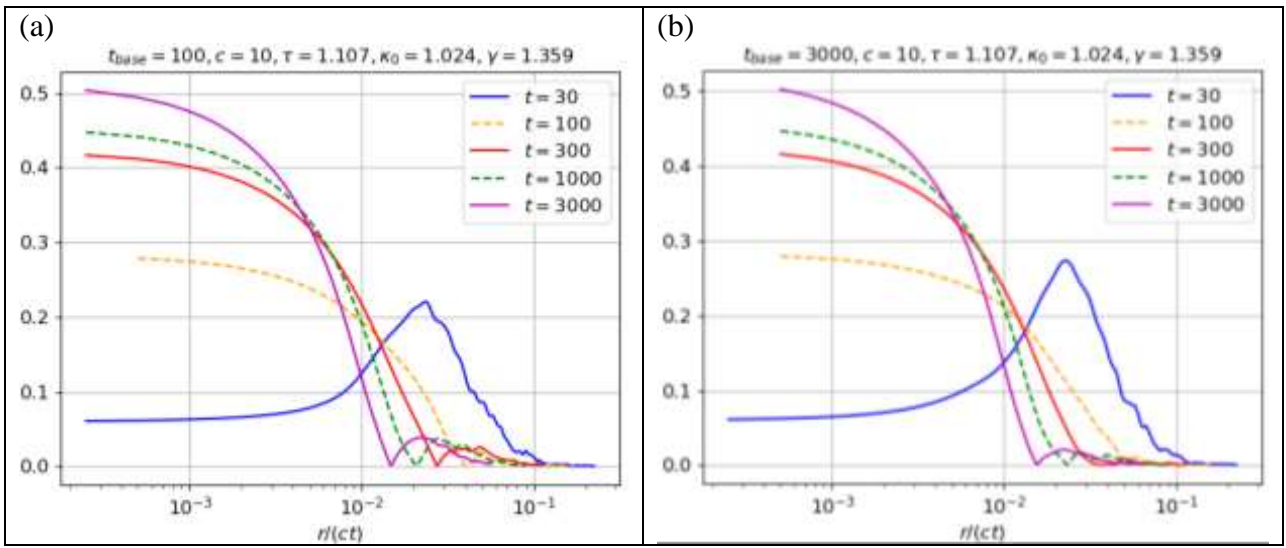

Figure 11. The ratio (17) when trajectory parameters in the self-similar solution are taken as reconstructed (Table 1, first line) and (a) $t_{base} = 100$, (b) $t_{base} = 3000$. Radial coordinate for each curve is limited to the region where the exact solution is not less than $10^{-3}$ of the value of this solution at $r = 0$.

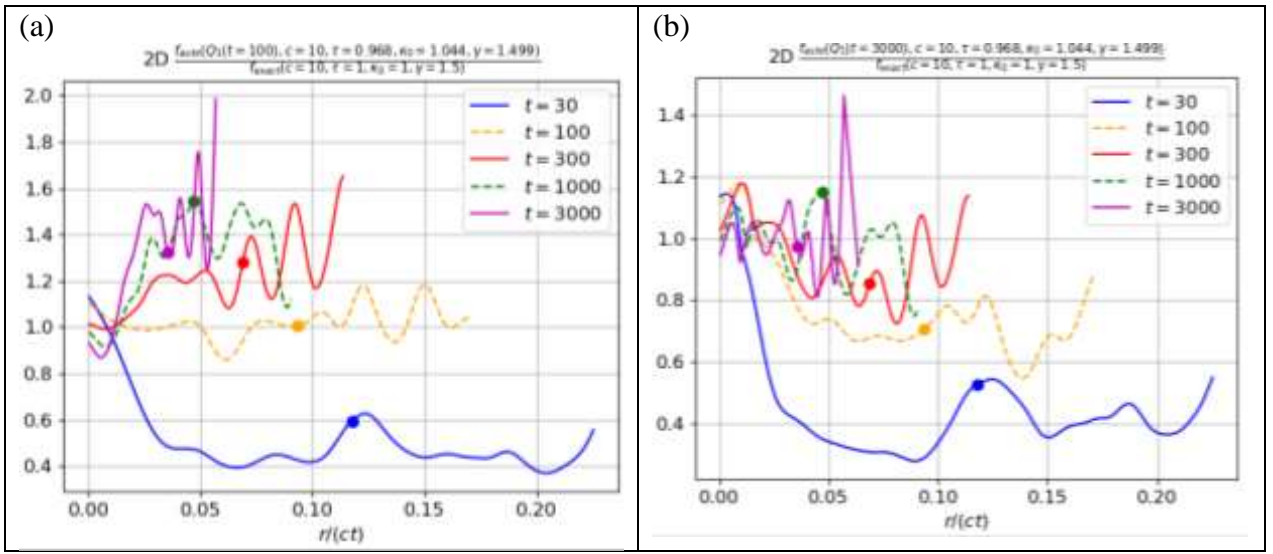

Figure 12. The same as in Figure 10 but for another values (second line in Table 1) of reconstructed trajectory parameters used in the self-similar solution.

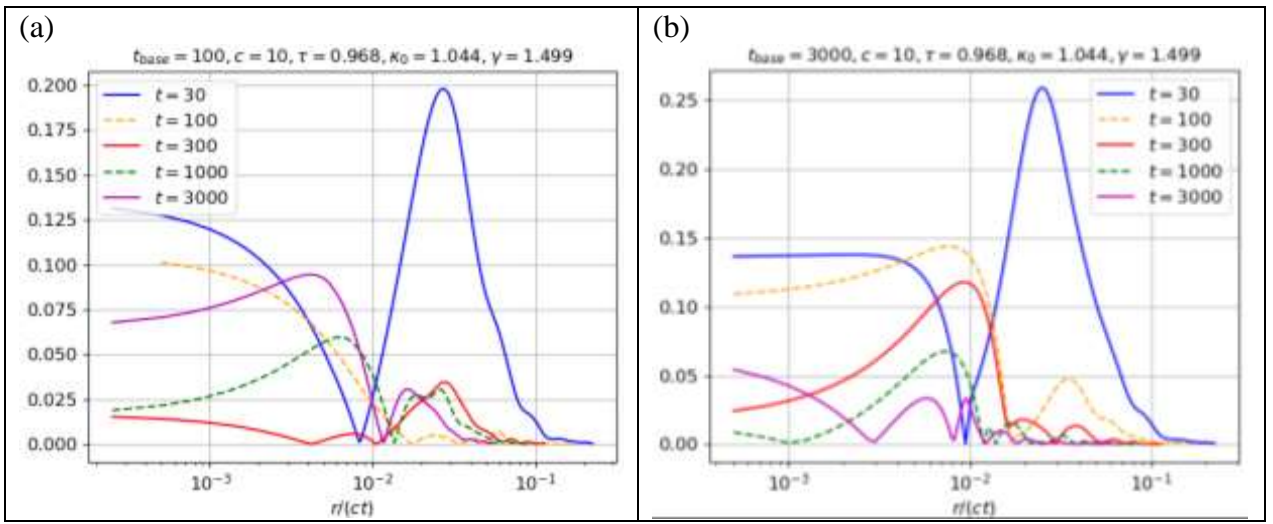

Figure 13. The same as in Figure 11 but for another values (second line in Table 1) of reconstructed trajectory parameters used in the self-similar solution.

Comparison of the results presented in Figures 10-13 shows that the accuracy of reconstructing the Green's function of the evolution of a migrant increases significantly, when the parameters of the migrant's trajectories are restored from a larger, but still not so many steps in the observed trajectories. The transition from the case of the first line in Table 1, that corresponds to ~$10^3$ steps in the trajectories, to ~$10^4$ steps (second line in Table 2), improves the accuracy of the spatial distribution of migrants by a factor of 2.

## 7. Conclusions

This work is an attempt to transfer to biology the methods developed in physics for formulating and solving kinetic equations in which the kernel of the integral operator in spatial coordinates is slowly decreasing and belongs to the class of Lévy distributions. Such an attempt is made in the frame of a synthetic diagnostics consisting in the generation of synthetic experimental data for trajectories of biological objects (migrants) and the subsequent reconstructions of the space-time evolution of the density of migrants. At first stage, the reconstruction procedure is applied to recover parameters of the probability density functions (PDFs) for the migrant's step length and the time of migrant's stay between runs using modest arrays of synthetic experimental data (moderate number of trajectories of migrants). The accuracy of this reconstruction is analyzed as a function of the volume of the used data. On the second stage, an approximate solution of the kinetic equation, obtained using previously developed method [27, 30], is tested against "true", synthetic experimental data for the PDF for the spatial density of migrants which start at certain point at certain time (i.e. against the Green's function of the corresponding integro-differential kinetic equation for the density of migrants in a broad space-time range).

The proposed algorithm is tested in the case which covers a wide class of time-dependent superdiffusion processes with a model power-law step-length PDF, which corresponds to "Lévy walks with rests" (cf. Figure 1 in the review paper [2]). The step-length PDF (1) is characterized by two parameters, the characteristic inverse length of the free path, $k_0$, and the power, $\gamma$, of the power-law PDF. These two parameters are related to the mean free path and the dispersion of the PDF. Note, however, that formally the variance of the Lévy-type PDF is infinite, and the formidable analysis of solution to kinetic equation is needed to evaluate the actual characteristics of a superdiffusion process. The waiting time PDF is chosen in a model exponential form with a single parameter, $\tau$, related to both the average time of migrant's stay between runs and the dispersion of the PDF. We considered here the case of the constant speed, c, of migrants. Thus the case of 4 parameters, with 3 free parameters among these, was considered to try the general algorithm proposed.

Comparison of approximate self-similar solutions for the parameters of step-length PDF and $\tau$, reconstructed from ~$10^3$ and ~$10^4$ steps in the observed trajectories, shows that the accuracy of reconstructing the Green's function of the evolution of a migrant increases significantly, when the parameters of the migrant's trajectories are restored from a larger, but still not so many steps in the observed trajectories.

**Acknowledgements**

This study was partially supported by the Russian Foundation for Basic Research (RFBR), grant No. 18-07-01269-a.

This work has been carried out using computing resources of the federal collective usage center Complex for Simulation and Data Processing for Mega-science Facilities at NRC "Kurchatov Institute", http://ckp.nrcki.ru/.

The authors are grateful to V.V. Voloshinov for help in using the Everest platform and the computing resources of the Center for Distributed Computing (http://distcomp.ru) of the Institute for Information Transmission Problems of the Russian Academy of Sciences, and A.P. Afanasyev for supporting cooperation between the NRC "Kurchatov Institute" and the above Center.